\documentclass[twocolumn,showpacs]{revtex4}

\usepackage{graphicx}
\usepackage{dcolumn}
\usepackage{bm}
\usepackage{hyperref}

\begin{document}

\title{Binary black hole mergers: large kicks for generic spin orientations}

\author{Wolfgang Tichy, Pedro Marronetti}
\affiliation{Department of Physics,
	Florida Atlantic University, 
	Boca Raton, FL 33431, USA}



%
\newcommand\be{\begin{equation}}
\newcommand\ba{\begin{eqnarray}}

\newcommand\ee{\end{equation}}
\newcommand\ea{\end{eqnarray}}
\newcommand\p{{\partial}}
\newcommand\remove{{{\bf{THIS FIG. OR EQS. COULD BE REMOVED}}}}
%

\begin{abstract}

We present results from several simulations of equal mass black
holes with spin. The spin magnitudes are $S/m^2=0.8$ in all cases,
but we vary the spin orientations arbitrarily, in and outside 
the orbital plane. We find that in all but one
case the final merged black hole acquires a kick of more than
1000~km/s, indicating that kicks of this magnitude are likely to
be generic and should be expected for mergers with general spin 
orientations. The maximum kick velocity we find is 2500~km/s
and occurs for initial spins which are anti-aligned in
the initial orbital plane.

\end{abstract}

\maketitle

\section{Introduction}

Currently several laser-interferometric gravitational wave detectors such as 
TAMA \cite{TAMA_web}, LIGO~\cite{LIGO_web} and GEO~\cite{GEO_web} are 
already operating, while several others are in the
planning or construction phase~\cite{Schutz99}. One of the most promising
sources for these detectors are the inspirals and mergers of binary black
holes.
The gravitational waves radiated during the inspiral and merger of two black
holes carry energy, linear momentum and angular momentum fluxes.
The radiated energy is positive
and thus the system will always lose energy. Except in the case of certain
highly symmetric situations, such as head on collisions of equal mass
non-spinning black holes, the system usually also loses angular momentum.
Generic systems with either unequal masses
or spins with no particular alignment will also radiate linear momentum,
resulting in a non-zero residual velocity (also known as recoil or ``kick") 
of the final black hole. The magnitude of this recoil
is important in a variety of astrophysical scenarios, such as
the cosmological evolution of supermassive
black holes~\cite{Merritt:2004xa,Boylan-Kolchin:2004tf,Haiman:2004ve,
Madau:2004st, Yoo:2004ze,Volonteri:2005pn,Libeskind:2005eh,Micic:2005gj}
or the growth and retention of intermediate-mass black holes in dense
stellar clusters~\cite{Miller2002a,Miller2002b,Mouri2002a,Mouri2002b,
Gultekin:2004pm,Gultekin:2005fd,O'Leary:2005tb}.
For a binary in almost circular orbit, the direction of the instantaneous
linear momentum flux rotates in the orbital plane with the angular
velocity of the system. Thus, when the binary goes through one orbital
period, the average linear momentum flux will be close to zero.
The only net effect comes from the fact that the inspiral orbits
are not perfect circles. Most of the kick is thus accumulated
during the last orbit and subsequent plunge of the two holes,
when the motion of the two holes is no longer quasi-circular and
the averaged linear momentum flux is much larger.
Several analytical estimates of the kick velocity
have been published in recent
years~\cite{Wiseman:1992dv,Favata:2004wz,Blanchet:2005rj,Damour:2006tr}.
All these estimates have been derived
using approximations (such as post-Newtonian theory) which
break down during the last orbit and the subsequent merger.
Hence these calculations contain a high degree of uncertainty.
In addition, there are calculations which employ the close
limit approximation~\cite{Sopuerta:2006wj, Sopuerta:2006et}.
However, these calculations can only model well the plunge and
miss contributions from the last orbit.
Furthermore, results from several numerical simulations have been
published~\cite{Herrmann:2006ks,Baker:2006vn, Gonzalez:2006md, 
Herrmann:2007ac, Campanelli:2007ew, Gonzalez:2007hi,
Campanelli:2007cg,Baker:2007gi}
recently.
The resulting maximum kick velocities found so far are:
175~km/s for non-spinning unequal mass black holes~\cite{Gonzalez:2006md},
and 2500~km/s for equal mass black holes with anti-aligned 
spins $S/m^2 \approx \pm 0.8$ in the orbital plane~\cite{Gonzalez:2007hi}.
The latter value is so high that the final black hole
would escape e.g. both from dwarf elliptical and spheroidal
galaxies (with typical escapes velocities of below 300~km/s) 
and from giant elliptical galaxies (2000~km/s)~\cite{Merritt:2004xa}. 
Extrapolations from simulations
of black hole binaries with spins in the orbital plane
predict velocities as high as 4000~km/s \cite{Campanelli:2007cg}.
The question thus arises if such high kick velocities 
are generic for spinning black holes, or if the results of 2500~km/s 
or even 4000~km/s arise only if the initial spins are exactly anti-aligned
and confined to the initial orbital plane.
To answer this question we have
performed several simulations of black hole binaries on initially
approximately circular orbits with spins in arbitrary directions.
We find that except for one case the kick velocities are well
above 1000~km/s. Thus, kicks of this magnitude seem
to be a generic feature of black hole mergers with spin around
or above $S/m^2 \approx 0.8$

\section{Numerical Techniques}
\label{techniques}

\begin{figure}
\includegraphics[scale=0.33]{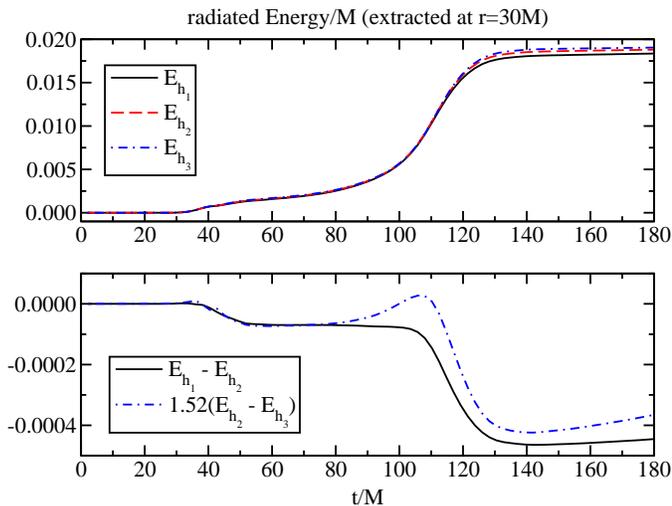}
\caption{\label{Energy_mmz}
This plot shows the energy radiated during
the merger of a binary where both initial spins ($S/m^2 =0.8$) are
anti-aligned to the orbital angular momentum.
The upper panel shows the energy radiated for the resolutions
$h_1=M/56.9$, $h_2=M/61.6$, $h_3=M/66.4$. In the lower panel
we show the differences scaled for fourth order convergence.
}
\end{figure}

\begin{figure}
\includegraphics[scale=0.33]{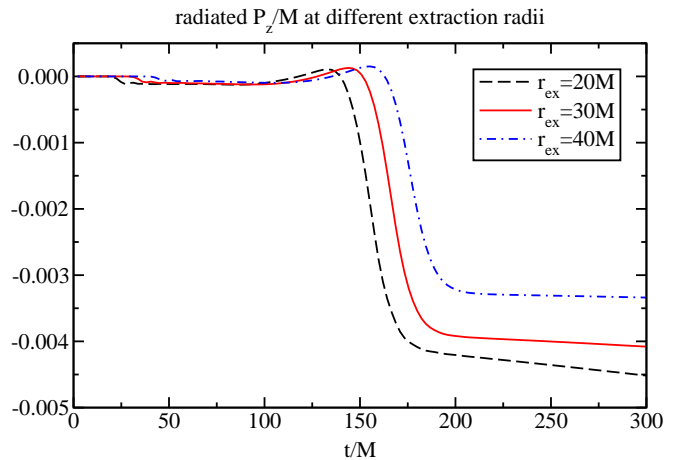}
\caption{\label{Smmxpmy_Pz_rex}
The z-component of the radiated momentum for the -1,135/-1,-135 run.
The initial spins are perpendicular to each other and lie in the initial
orbital plane. This plot shows that on our relatively
small grids the location of the extraction radius plays a
big role and is the main contributor to the error of $\pm$20\% quoted here. 
}
\end{figure}

Our simulations are performed using 
the ``moving punctures" method \cite{Campanelli:2005dd,Baker:2005vv}
with the BAM code~\cite{Bruegmann:2003aw,Bruegmann:2006at}
which allows us to use moving nested refinement boxes.
We use 10 levels of 2:1 refinements. The outer boundaries
are located $240M$ away from the initial center of mass
($M$ being the sum of the initial black hole masses),
and our resolution ranges between $9M$ on the outermost box
to $M/56.9$ near the black holes. Since our simulations
have no particular symmetries we cannot use the usual memory
saving techniques and simulate only one quadrant of the numerical
domain. Thus our simulations take four times more memory
and time than e.g. the ones reported in~\cite{Marronetti:2007ya}.
This makes standard convergence tests very costly. Thus,
the goal of this work is merely to give kick velocity estimates
which can be used in astrophysics,
instead of extremely accurate results.
For the same reason, we present a convergence test for only one
particular case in order to establish that our numerical
setup gives results that are in the convergent regime.
In Fig.~\ref{Energy_mmz} we show the energy radiated during
the merger of a binary where both initial spins 
have magnitude $S/m^2 =0.8$ and are
anti-aligned to the orbital angular momentum.
The lower panel of this plot demonstrates that the resolutions
$h_1$, $h_2$ and $h_3$ are in the convergent regime. 
All other results which are given below, have been obtained
using the lowest resolution $h_1=M/56.9$. From
this study it is also possible to obtain error bars due
to the finite differencing scheme used here. These error
bars are of the order of a few percent.
Notice however, that there are other sources of errors,
the biggest of which is the extraction of the radiated energy
and momentum at a finite radius. This radius should ideally be far
from the sources. However, the outer regions of our grid 
are not very well resolved, since the highly refined boxes do not 
extend far from each black hole. We find that the best compromise 
for our grid structure is to extract at $r_{ex}\sim 30M$.
Figure \ref{Smmxpmy_Pz_rex} shows the radiated momentum
in the $z$-direction extracted at different radii.
In this example both initial spins are perpendicular to each other and
lie in the initial orbital plane.
We use the difference between these results at different radii
as error bars for the kick velocities presented below.

\section{Results}
\label{results}

In order to gauge possible kick sizes we have performed several
simulations with different spin orientations. In each case
the spin magnitude of each black hole is $S/m^2=0.8$
and the initial orbital angular momentum is along the $z$-axis.
The magnitude of the orbital angular momentum is chosen
such that the orbits are circular according to post-Newtonian
approximations~\cite{Kidder:1995zr,Marronetti:2007ya}.
The bare mass parameter $m_b$ of each puncture is chosen such
that the ADM mass $m$ measured at each puncture is approximately $0.5M$.
Because of these approximations, it is clear that the orbits
will have small ellipticities. However, since we are just
interested in obtaining a sample of possible kick velocities
for different initial conditions such imperfections are not critical.

We label each simulation using seven numbers: [$q$, $D$, $S_1/m^2$,
$90^\circ-\theta_1$, $\phi_1$, $S_2/m^2$, $90^\circ-\theta_2$,
$\phi_2$]. Defining the larger black hole as $\# 2$, we set 
$q \equiv m_2/m_1 \geq 1$. $D$ and $S_A/m^2$ represent the holes' 
coordinate separation and the spin magnitude of hole $A$ ($A=1,2$). The 
angles $\theta_A$ and $\phi_A$ (given in degrees) determine
the direction of each spin by their polar angles with respect to
a coordinate system ($\hat{x}$, $\hat{y}$, $\hat{z}$), where 
$\hat{z}$ is the unit vector in the direction of the binary's
initial orbital angular momentum, $\hat{x}$ is the unit vector in the
direction of the initial linear momentum of the larger hole and 
$\hat{y} = \hat{z} \wedge \hat{x}$ \footnote{In this paper, we 
constructed our initial data sets such that the numerical grid
coordinate system ($x$, $y$, $z$) coincides with
($\hat{x}$, $\hat{y}$, $\hat{z}$)}.
Since our runs all have equal masses and spin magnitudes, they are
all of the form 
[$1$,$6M$,$0.8$,$90^\circ-\theta_1$,$\phi_1$,$0.8$,$90^\circ-\theta_2$,$\phi_2$].
The runs in Table~\ref{pars_kicks} are thus labeled by the angles alone.

From Table~\ref{pars_kicks} we see that almost all
initial spin orientations lead to large kick velocities.
The only exception is when both initial black hole spins
point in exactly the same direction (see run -31,0/-31,0). 
In this special case no net linear momentum is radiated
(in agreement with post-Newtonian predictions~\cite{Kidder:1995zr}).
We find that the largest kick occurs for 
anti-aligned spins that are perpendicular to the initial
angular momentum $J^{ADM}_{\infty}$.
However, we also find substantial kicks
for other initial spin orientations, and in each case
the kick is well above 500~km/s. This means that kicks
of this magnitude should be expected for generic black hole
mergers with spins of magnitude of $S/m^2 = 0.8$
or above. 

\begin{figure}
\rotatebox{-90}{\includegraphics[scale=0.37,clip=true]{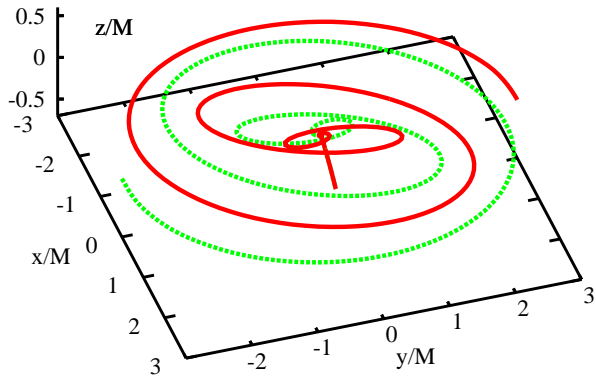}}
\caption{\label{Sxz-tracks}
This plot shows the tracks of the black hole centers for the
-1,180/89,180 run.
Here one of the initial spins is in the initial orbital plane, 
and the other one is perpendicular to the orbital plane.
One can see that the orbits do not stay in one plane.
}
\end{figure}

This result can be explained in the following way:
it is known that the black hole spins significantly influence the
dynamics of the two black holes. For example, the merger can be delayed if
both spins are aligned with the orbital angular
momentum~\cite{Campanelli:2006uy}. Or the orbits can be non-planar as in the
case of a test mass in orbit around a spinning Kerr black hole
\cite{Campanelli:2006fy}. Since the
linear momentum radiated depends on the orbit shape, the
kick velocity also depends strongly on spin.
In fact, if the spins are not parallel to
the orbital angular momentum, the spins and the orbital
angular momentum both precess so that the resulting orbits do not stay in
one plane and become much more complex than the familiar Newtonian
circular orbits. Since the instantaneous
linear momentum flux changes direction in a less regular way
than for circular orbits, one expects that the
averaged linear momentum flux is significantly larger than for
quasi-circular orbits. As an example of such a more complicated
orbit see Fig. \ref{Sxz-tracks} which shows the orbits of both
black hole centers 
for the case where one of the spins starts out in the initial orbital
plane and the other one perpendicular to this plane.
We can clearly see that the orbits do not stay in one plane,
and that the kick moves the final merged black hole
downward out of the initial orbital plane.

From Figs. \ref{Energy_mmz} and \ref{Smmxpmy_Pz_rex} we can
see that the bulk of the radiated energy and momentum
is emitted within a time of about $40M$. This is also
the time during which the maximum amplitude of
the gravitational waves generated by the merger pass
through the detector. I.e. almost the entire
kick is accumulated during the final merger phase.
Table~\ref{pars_kicks} gives estimates for the mass and spin
of the final black hole.

\begin{widetext}

\begin{table}
{\small
\begin{tabular}{c|c|c|c|c|c|c|c|c}
Run			&0,90/0,-90 & -1,135/-1,-135 & 45,0/-45,180 & 44,180/-46,180 & -31,0/-31,0 & -1,180/89,180 & -1,150/89,151 & -1,120/89,119	\\
\hline
$M^{ADM}_{\infty}/M$	& 0.985  & 0.985  & 0.985  & 0.985  & 0.985  & 0.984  & 0.984  & 0.984	\\
$J^{ADM}_{\infty}/M^2$& 0.842	 & 0.882  & 0.839  & 0.881  & 0.756  & 1.016  & 1.016  & 1.016	\\
 $m/M$			& 0.50 	 & 0.50   & 0.50   & 0.50   & 0.50   & 0.50   & 0.50   & 0.50	\\
 $m_b/M$	 	& 0.30 	 & 0.30   & 0.30   & 0.30   & 0.30   & 0.30   & 0.30   & 0.30	\\
 $D/M$			& 6.000	 & 6.000  & 6.000  & 6.000  & 6.000  & 6.000  & 6.000  & 6.000	\\
 $P/M$			& 0.140	 & 0.140  & 0.140  & 0.140  & 0.147  & 0.133  & 0.133  & 0.133	\\
$L/M^2$			& 0.842	 & 0.840  & 0.839  & 0.839  & 0.881  & 0.798  & 0.798  & 0.798	\\
$S_{1x}/m^2$		& 0	 &-0.566  & 0.566  &-0.576  & 0.684  &-0.800  &-0.693  &-0.400	\\
$S_{1y}/m^2$		& 0.800	 & 0.566  & 0	   & 0	    & 0	     & 0      & 0.400  & 0.693	\\
$S_{1z}/m^2$		& 0	 &-0.010  & 0.566  & 0.556  &-0.414  &-0.010  &-0.010  &-0.010	\\
$S_{2x}/m^2$		& 0	 &-0.566  &-0.566  &-0.556  & 0.684  &-0.010  &-0.009  &-0.005	\\
$S_{2y}/m^2$		&-0.800  &-0.566  & 0	   & 0	    & 0	     & 0      & 0.005  & 0.009	\\
$S_{2z}/m^2$		& 0	 &-0.010  &-0.566  &-0.576  &-0.414  & 0.800  & 0.800  & 0.800	\\
\hline
$M_f/M$			& 0.95	 & 0.95   & 0.95   & 0.952  & 0.96   & 0.94   & 0.94   & 0.94	\\
$J_f/M_f^2$		& 0.67	 & 0.72	  & 0.68   & 0.73   & 0.64   & 0.81   & 0.80   & 0.80	\\
$v$~(km/s)  		& 2500   & 1350   & 580    & 1350   & 0      & 1850   & 2100   & 1950	\\
			&$\pm500$&$\pm250$&$\pm150$&$\pm250$&$\pm50$ &$\pm350$&$\pm450$&$\pm400$\\
\end{tabular}
}
\caption{\label{pars_kicks}
Initial parameters and resulting kick velocities. The runs are labeled
by the spin orientation angles, which for each hole are the
angle between the spin and the initial orbital plane and the angle
between the projection of the spin into the orbital plane
and the initial linear momentum of hole $2$.
Here $m_b$ is the bare mass parameter of each
puncture and $m$ the ADM mass measured at each puncture. 
The black holes have coordinate separation $D$, with puncture 
locations $(0, \pm D/2 ,0)$ and linear momenta $(\mp P, 0, 0)$,
so that the initial orbital angular momentum is in the $z$-direction. 
The spins have a magnitude of $S/m^2 =0.8$ and
point in the various directions listed here, so that
the total angular momentum is not necessarily along the $z$-direction.
We also list the estimates for the initial ADM 
mass $M^{ADM}_{\infty}$ and ADM angular momentum $J^{ADM}_{\infty}$.
Furthermore, we give the final black hole mass 
$M_f$ and angular momentum $J_f$, as well as the kick velocities $v$
for the different cases.
}
\end{table}

\end{widetext}

\section{Discussion}
\label{discussion}

We have performed simulations of equal mass black holes
with initial spins of magnitude of $S/m^2=0.8$.
We have tested several initial spin orientations
and find that all but one leads to kick velocities well
above 1000~km/s. Thus kicks between 1000~km/s and 2000~km/s seem
to be a generic feature for spins of magnitude $0.8$ with
general orientations. Higher spins will lead to even larger kicks.
This means that the final black hole
will likely escape from dwarf elliptical and spheroidal
galaxies (with escape velocities below 300~km/s), and possibly even from
giant elliptical galaxies 
(with escape velocities around 2000~km/s)~\cite{Merritt:2004xa}.
This result also has important implications for
a variety of astrophysical scenarios, such as
the cosmological evolution of supermassive
black holes or the growth and retention of intermediate-mass 
black holes in dense stellar clusters.
Note, however, that a more extensive parameter search should
be carried out before drawing any definitive conclusions.
Furthermore, notice that all our simulations are
for the equal mass case. It is clear that the kick velocity
will depend on the mass ratio 
(for two proposed models see~\cite{Campanelli:2007ew,Baker:2007gi}),
and we expect smaller kicks for the unequal mass case.
The probability for the large kicks obtained here might be quite low,
if kick velocities drop sufficiently fast for mass ratios 
different from one.

Almost the entire kick is accumulated during the final merger phase,
meaning that small changes in the initial conditions
are likely to lead to quite different kick directions and magnitudes.
The reason is that a small change in the initial velocities
or separation (and also accumulated numerical errors in the orbital phase),
will significantly alter the merger time~\cite{Bruegmann:2006at},
and thus alter the spin orientations just before merger. 
We therefore expect different trajectories just before the
merger which then will lead to different final kicks.
Note, however, that this sensitivity to initial conditions
does not change our main conclusion that generic spin
orientations can lead to very large kicks in the equal mass case.

\begin{acknowledgments}

It is a pleasure to thank B. Br\"ugmann, J. Gonz\'alez, 
M. Hannam, S. Husa, and U. Sperhake for helpful comments and
for implementing wave extraction tools in the BAM code.
This work was supported by NSF grant PHY-0555644.
We also acknowledge partial support by the
National Computational Science Alliance under
Grants PHY060021P and PHY060040T.
Simulations were also performed at the
Charles E. Schmidt College of Science computer cluster Boca 5.

\end{acknowledgments}




\bibliography{references}

\end{document}